\documentclass{article}
\usepackage{amssymb}
\usepackage{epsfig}
\usepackage{makeidx}
\usepackage{graphicx}
\usepackage{multicol}
\usepackage{amsmath}
\usepackage{amsfonts}

\begin{document}

\title{\textbf{On inverse problem of  waves identification by measurements at one point vicinity}}
\author{S.Leble$^{1,2}$, I.Vereshchagina$^{2}$ \\ \small 1. Gdansk University of Technology, Faculty of Applied
Physics and Mathematics,\\  \small ul. Narutowicza 11/12, 80-233 Gdansk,
Poland, \\ \small 2. Immanuel Kant Baltic Federal University, ul.A. Nevskogo, 14, 236041, Kaliningrad,  Russia.
\\ leble@mif.pg.gda.pl,ver.is@mail.ru}

\maketitle
\begin{abstract}
A problem of a wave identification is formulated. An example is considered in conditions of one-dimensional Cauchy problem for conventional string equation in matrix form and its  inhomogeneous two-component version. The acoustic and electromagnetic problems are discussed within the restrictions outlined. The projecting operator technique is used to split the solution space and analyze input of a wave monitoring in vicinity of an observation   point. The solution space is supplied by $L_2$ norm via the problem conservation law; its finite-dimensional analog is used as a measure of a given mode presence and information about form.  The
algorithm of the problem solution is presented in terms of appropriate regularization to reconstruct an incoming pulses origin. The dissipation and entropy mode  account in the problem of acoustic waves extraction   is also discussed in terms of correspondent projecting technique.

\end{abstract}

\section{Introduction}

A problem of a wave mode identification has interdisciplinary character, it is important, for example
in physics of atmosphere, where superposition of acoustical, gravity and planetary waves occur \cite{pedlosky}. In planetary range 
the periods of Rossby and Poincare waves are of the same scale, hence their separation and estimation of contributions is complicated. The situation is even more complicated  in plasma physics for additional specific  branches of waves coexists with ones for neutral gas. An important problem of a specific mode  source localization is also typical inverse problem that is generally ill-posed \cite{Lavr}.

In geophysics the wave field diagnostics generally needs many observations that cover a space enough for wave length estimation. It is rather expensive and not very feasibly.
Thinking about the novel alternative approach \cite{LeKar} we suggest use measurements in a vicinity of a point but many-components observations with aid of the projecting operators technique built in this article, such that fit subspaces of  specific waves.

We would start with an instructive  example of 1D wave equation and correspondent Cauchy problem (Sec. \ref{CP}), see e.g. \cite{RM,Lavr1}. 
Naturally a dispersion or dissipation complicate the situation but may be overcome \cite{L}. Weak inhomogeneities of propagation media may be also effectively included  in similar manner \cite{LeVer}.

  There are lot of important problems  of theoretic physics the same level of description: for electrodynamics see \cite{Kins1}, \cite{Kins2},\cite{KL} for acoustic \cite{Pere},\cite{Pere2}, and Tollmienn-Schlichting waves \cite{PerLe}.
that may be directly formulated as the  system of equations so the vector description (in electrodynamics it is (E,B)) have direct physical sense. Mention however that the directed waves correspond to so-called hybrid variables with appropriate initial or  boundary conditions.

More complicated 3D problems need more advanced constructions and geometry (ring or sphere in geophysics \cite{LeKar} - e.g.) impact  may lead to very non trivial  generalization of the technique and algorithm as well as norm of the appropriate spaces construction.

Our present  study is  focused on the simplest 1+1 case, that include one space and one time coordinate. We restrict ourselves by uniform medium mainly, but would like to touch the base of projecting operator technique and the estimation of diagnostics quality in a finite number of measurements realization in terms of physically reasonable Banach space. Hence we could estimate the diagnostics errors and, therefore, quality of position of a wave source estimation.
We fix our attention on the minimal version of the theory to show the main idea of a wave diagnostics, may be most characteristic for the opposite propagating waves. In this 1+1 case the wave type (polarization) is linked to direction of propagation, that allow to formulate the whole algorithm of some inverse problem solution in the following form:

1) The refrmulation of a problem that fix eigen subspaces of evolution operator. 

2)  Projecting to subspaces and their  weight evaluation in an appropriate physical norms for functional and finite-dimensional spaces.

3) Time arrival  and the wave form estimation for a given number of measurements and spline order choice.

4) Estimation of a distance to the area of initialization within the prescribed error limits.

5) Investigation of stability in terms of explicit solution form, reconstructed by the finite points number  data.
Mention that the last point relates to analytical continuation problem \cite{Lavr}. 

Such algorithm is realized in the Sec.\ref{CP} for the Cauchy problem for homogeneous string equation example and developed in the Sec \ref{IM} for more general system and weak inhomogeneity account.
A development of the theory with dissipation  and Entropy mode account in Sec. \ref{DA}.

 \section{String equation in vector form. Cauchy problem in terms of projectors.}
 \label{CP}
 
The conventional Cauchy problem for the wave equation contains two initial condition, including time derivative,  needs a measurement  of the derivative hence its physical version includes few points in the vicinity of the point of observation and estimation of diagnostics error in an appropriate space. The conventional Cauchy problem for the 1+1 wave equation
\begin{equation}
	u_{tt}-c^2u_{xx}=0
\end{equation}
 with initial data

$ u(x,0)=\phi(x),\quad v(x,0)=\psi(x)$.

 has the matrix  representation in terms of a vector 
 
\begin{equation}\label{psi}
	\psi^T=(cu_x=v,u_t=w), D=\frac{\partial}{\partial x}
\end{equation}
 and evolution operator $L$, that appear in the system
\begin{equation}\label{uw}
\begin{array}{c}
v_t-cw_x=0,\\
	w_{t}-cv_{x}=0.
	\end{array}
\end{equation}
\begin{equation}
\psi_t=cD(
\begin{array}{c}
 w\\
	v
	\end{array})= cD
(
\begin{array}{cc}
 	0&1\\
			1&0
	\end{array})\psi	=L\psi,	 
 \end{equation}
that may solved with projecting operator \cite{L} technique. The result for the system (\ref{uw}) is almost obvious
\begin{equation}
	P_{\pm}=\frac{1}{2}(
\begin{array}{cc}
 	1&\pm 1\\
			\pm1&1
	\end{array}). 	
\end{equation}
The identity
\begin{equation}
	(P_++P_-)\psi=\psi
\end{equation}
in terms of 
\begin{equation}
	P_+\psi=(\begin{array}{c}
 \Pi\\
	\Pi
	\end{array})
\end{equation}
and
\begin{equation}
	P_-\psi=(\begin{array}{c}
 \Lambda\\
	-\Lambda
	\end{array})
\end{equation}
reads
\begin{equation}
	\begin{array}{c}
 \Pi=\frac{1}{2}(v+w),\\
	\Lambda=\frac{1}{2}(v-w).
	\end{array}
\end{equation}
For details and generalization see, however \cite{LeVer}.

Even in this simlest case the physical content is nontrivial because the right (left) wave is of a hybrid form (see \cite{L,Kins1}), recall that the definition of the variables $v,w$ contants derivatives \eqref{psi},  therefore $\Pi=\frac{1}{2}(cu_x+,u_t)$. As it is known, the derivative evaluation in a finite-difference context is ill=posed and  needs a regularisation \cite{Lavr}.

The projecting  separates the space $\Psi$ into direct sum of subspaces 
$$
\Psi=\Psi_+\oplus\Psi_-,(\begin{array}{c}
 \Pi\\
	\Pi
	\end{array})\in\Psi_+ 
$$

Applying directly to the (\ref{uw}) yields  
\begin{equation}
	(P_{\pm}\psi)_t=cDP_{\pm}(
\begin{array}{cc}
 	0&1\\
			1&0
	\end{array})\psi
\end{equation}
or
\begin{equation}\label{LPe}
	\begin{array}{c}
\Pi_t+c\Pi_x=0,\\
\Lambda_t-c\Lambda_x=0.
	\end{array}
\end{equation}

The technique is a convenient for analysis of the problem of diagnostics and its quality estimation because the space of solution splits to ones with more simple evolution. For the  one-dimensional string case only right and left waves are taken into account. 

Let a right wave $\Pi$ arrives to a point of observation, say $x=0$, giving in a measurement of the vector $\psi$ components. Action of the "left" projector $P_-$ to such vector gives zero, if the function $\Pi$ is exact solution of the equation (\ref{LPe}). 

Suppose the vector space of solutions is attributed by a norm, $\psi\in B$, hence  $B$ being a Banach space. 
In a general diagnostic exposition  introduce a normalized solutions $\lambda,\pi$ of more complicated situation when a  disturbance of the string $\alpha \pi+\beta\lambda$ may arrive at $x=0$ from both sides simultaneously, then action of the projectors $P_{\pm}$  cuts one of the waves with the result, for example
\begin{equation}
	P_+\psi=P_+[\alpha(\begin{array}{c}
 \pi\\
	\pi
	\end{array})+\beta(\begin{array}{c}
 \lambda\\
	-\lambda
	\end{array})]=\alpha(\begin{array}{c}
 \pi\\
	\pi
	\end{array})
\end{equation}
 Unfortunately a real measurement gives  the information about $\psi$ components with errors and only in a finite number of time points. So we should estimate a distance between a representative observation and a space of, say,   right waves.  

Let us introduce a norm in the vector space of solutions, decaying at infinities exponentially. It is directly verified by \eqref{LPe}
\begin{equation}\label{cl}
	[\int_{-\infty}^{\infty}(\Lambda^2+\Pi^2)dx]_t=0,
\end{equation}
therefore it is convenient to introduce the norm via this conservation law \eqref{cl}
\begin{equation}\label{norm}
	||\psi||^2=\int_{-\infty}^{\infty}(\Lambda^2+\Pi^2)dx,
\end{equation}
or in terms of original components 
\begin{equation}
	||\psi||^2=\int_{-\infty}^{\infty}(\frac{1}{2}(v^2+w^2))dx,
	\end{equation}
	because $\Lambda^2+\Pi^2=	\frac{1}{4}[(v+w)^2+(v-w)^2]$, the integral is proportional to energy density of a string.
	
	Such way a Banach space $\Psi$ is introduced 
to estimate distances in the space of $\psi \in \Psi$, with the normalized $||\lambda||=||\pi||=1.$

Given a sequence of times $t_i, i=1,...n$ generates a sequence $\phi(0,t_i)$, by measurements, which form n-dimensional vector  $\phi\in R^n$  with a norm 
\begin{equation}\label{normn}
	||\phi||_n^2=\sum_{i=1}^n(\phi_1^2(0,t_i)+\phi_2^2(0,t_i)),
\end{equation}
 that should determine a closest vector solution  $\psi_+(x-ct)$ via the functional minimum
\begin{equation}\label{F}
I=||\psi_{+}-P_+\phi(0,t_i)||_n,
\end{equation}
where $\psi_{i+}=\psi_+(0-ct_i).$
We treat this condition as variational principle 
$$
	min_{\psi_+\in \Psi_+} I,
$$ 
in a   n-dimensional space with the norm (\ref{normn}) correspondent to $L_2$  (\ref{norm}) with the Euler equations
\begin{equation}
	\frac{\partial I}{\partial \psi_{i+}}=0.
\end{equation}
Let us recall (see (\ref{psi})) that the vector of observation data have the components $\phi_1=c\frac{\Delta  u}{\Delta x} \approx cu_x, c \frac{\Delta u}{\Delta t} \approx cu_t), $ that are obtained by measurements of the string function $u(x,t)$ in adjacent points of time and in $x=0$ and in $x=\Delta x$. For example, if the sequence of components coincide ($\phi_1(t_i)=\phi_2(t_i)$) the action of the left projector $P_-$
\begin{equation}
	P_{-}\phi=\frac{1}{2}(
\begin{array}{cc}
 	1&- 1\\
			-1&1
	\end{array})(\begin{array}{c}
 	\phi_1\\
			\phi_1
	\end{array})
\end{equation}
	gives zero.
In reality the errors of measurements do not give zero identically even if the wave is purely right. The deviation of the real data from the ideal may be characterized by
    \begin{equation}
  ||\phi_{+}-P_+\phi(0,t_i)||_n=\delta.
\end{equation}
the correspondent calibration may be performed by the specially organized experiment.
   
An admixture of a left wave in a superposition of the both may be noticed if the $||\phi-P_+\phi||_n$ exceeds the typical error.  If noticed the left wave reconstruction of the second step is made by the second projector $P_-$.

Final form of the reconstruction is a standard procedure of spline or other inverse problem in a chosen a priori physically reasonable space \cite{Lavr}. Note, that arrival time needs a priori information about "zero" time event.
An instability from velocity value errors should also be taken into account, error from arrival time grows with ct.

\section{General hyperbolic problem. Inhomogeneity account}
\label{IM}

Here we admit a inhomogeneity of the medium of propagation in more natural frame of the system for a couple of directly measured variables $u,v$.  In acoustics it would be a hydrodynamic velocity and pressure \cite{Pere}, in electromagnetism - electric and magnetic fields \cite{Kins3}. The tsunami problem is studied in \cite{RM} within very similar framework.  Such system and initial problem data for a vector $\psi^T=(u,v)$
do not contain a derivative in space, so its diagnostics needs the only point time sequence measurements of both vector components.  
 
Consider the initial problem for the  system
\begin{equation}\label{1}
\frac{\partial u(x,t)}{\partial t }
- \epsilon b(x)\frac{\partial v(x,t)}{\partial x}=0,
\end{equation}
\begin{equation}\label{2}
\frac{\partial v(x,t)}{\partial t }
- \epsilon c(x)\frac{\partial u(x,t)}{\partial x}=0.
\end{equation}
for $u(x,t,v(x,t) \in C^1, x\in(-\infty,\infty),t\geq 0,\, u(x,0)=\phi(x),v(x,o)=\psi(x).$
 We  introduce a small parameters $\epsilon$ , to characterize the initial conditions 
\begin{equation}
max \frac{\partial \phi(x)}{\partial x } = \epsilon  << 1 ,
\quad max \frac{\partial \psi(x)}{\partial x } = \epsilon << 1,
\end{equation}
The (weak) inhomogeneity is described by the system coefficients $c,d$ dependence on x. A dependence on the small parameter $\epsilon$ is implied and skipped in this text (see details in \cite{LeVer}). 

The problem is reformulated in terms of directed waves. The projecting operators   in this case are calculated via the basic relation   for the projection subspaces that are derived directly from equations (\ref{1},\ref{2}), in which evolution is fixed by the pseudodifferential spectral operators as expansion in $\epsilon$ \cite{LeVer}). In the first order, arriving at a supermatrix:
\begin{equation}\label{P12}
  P_{1,2}=\frac{1}{2} \left(
       \begin{array}{cc}
         1 & \pm M^{-1} \\
         \pm M & 1\\\
       \end{array}
     \right)
\end{equation}
where $f=\sqrt\frac{c}{b}, D=\frac{\partial  }{\partial x}$ and the operator valued matrix elements are expressed in terms of $M = D^{-1} f D$.

Now the evolution 
operator $L$ , in the same notations, is also supermatrix:
\begin{equation}
  L=\left(
       \begin{array}{cc}
         0 & \epsilon b(x)D\\
         \epsilon c(x)D& 0 
       \end{array}
     \right).
\end{equation}

To proceed in the theory we based on the commutation relation  $[P_{1,2},L]=0$ that is valid automatically in the case of constant coefficients $b,c$. For the x-dependent case of $b(x), c(x)$ the commutator $L$ and $P_1$ is equal to
\begin{equation}
[P_{1}, L]=\frac{\epsilon}{2}\left(
        \begin{array}{cc}
          M^{-1} c D - b D M & 0 \\
					         0 &  M b D - c D M^{-1} 
       \end{array}
     \right).
     \end{equation}

Condition that the commutator is zero can be written as
\begin{equation}
D^{-1}f^{'} b f=0,
\end{equation}
 or from the expression for f:
\begin{equation}
  c^{'}b -b^{'}c = 0.
\end{equation}
 It fix the case of complete reduction (diagonalization) of evolution operator. 
 
 As the further development of the method we suggest an approximate procedure (see e.g. \cite{PerLe}), generally treating the condition $[P_{1,2},L]=O(\epsilon).$
   
Using the  projecting operators  we reduce (\ref{1}) to a couple of equations that, in previous section we    traditionally name  ones for left and right waves, splitting the problem of evolution. The approximate splitting is achieved if one could neglect the commutators of $P_{1,2}$ and $L$. It is possible if the coefficients $b,c$ are of the zero order ($\cong O(1)$), while the order of the  derivative $(\frac{c}{b})'$ is of a higher order, e.g. $\cong O(\epsilon)$. It is guaranteed by the evolution operator dependence on $\epsilon$  and conditions of the spectral operators expansion domain. Acting by $P_1$ to the system \eqref{1},\eqref{2}
 \begin{equation}
\begin{array}{cc}
  (P_{1,2}\Psi)_t=P_{1,2}L\Psi, 
	\end{array}
	\end{equation}
or, approximately	
	\begin{equation}
	\begin{array}{cc}
  (P_{1,2}\Psi)_{t}=L(P_{1,2}\Psi),  
	 \end{array}
	\end{equation}
where 
	\begin{equation}
  P_{1}\Psi= 
		=\frac{1}{2} \left(
       \begin{array}{cc}
         \Pi  \\
         M \Pi\\
       \end{array}
     \right),
\end{equation}
		\begin{equation}
  P_{2}\Psi= 
		=\frac{1}{2} \left(
       \begin{array}{cc}
         \Lambda  \\
         - M \Lambda \\
       \end{array}
     \right).
\end{equation}
Reading the first lines of the relations yields	
\begin{equation}\label{pi}
	  \Pi=\frac{1}{2} (u + M^{-1} v),\\
			\end{equation}
and		
	\begin{equation}\label{la}
	  \Lambda=\frac{1}{2} (u - M^{-1} v).\\
			\end{equation}
This relation allows to establish the Cauchy problems for directed waves. 
\begin{equation}\label{pi}
	  \Pi(x,0)=\frac{1}{2} (\phi + M^{-1} \psi),\\
			\end{equation}
and		
	\begin{equation}\label{la}
	  \Lambda(x,0)=\frac{1}{2} (\phi - M^{-1}\psi).\\
			\end{equation}
			From the equations \eqref{pi},\eqref{la} one extracts:
			\begin{equation}\label{uv}
	\begin{array}{cc}
  u = \Pi+\Lambda,\\
	v = M(\Pi-\Lambda)
	 \end{array}
	\end{equation}

	Considering equations (25) ,(29) and the relation for the commutator $ P_1 L = L P_1 - [P_{1}, L] $ one obtains approximately:
		\begin{equation}\label{pir}
	  \Pi_{t}= -\sqrt{bc} \Pi_{x}, 
	 	\end{equation}
	That could be interpreted as the equation for the right wave.  	Similarly the equation for the left wave variable $\Lambda$ looks as follows
\begin{equation}\label{lar}
	  \Lambda_{t}= \sqrt{bc}\Lambda_{x}.
 	\end{equation}
	 Solving the first order equations by method of characteristics gives $u, v$ by the relation \eqref{uv}; formally the system coincides with \eqref{LPe} but velocity of propagation and coefficients in \eqref{uv} are  functions depending on coordinate.

The problems of diagnostics and reconstruction is formulated within the scheme of the previous section, is  based on the algorithm described in introduction. 
 The norm based on left/right waves variables representation (\ref{norm}) may be reformulated from the conservation law
\begin{equation}\label{con2}
	  [\int_{-\infty}^{\infty}(b^{-1}u^2+c^{-1}v^2))dx]_{t}= 0,
 	\end{equation}
for integrable integrand (e.g. $b>0,c>0$).
The correspondent functional  is again invariant with respect to time shift.  The functional (\ref{F}) has the same form but the projector in it  is modified as in (\ref{P12}). The perturbation, again settled at time $t=0$ localized as prescribed the Banach space $\Psi$, propagates along characteristics for given inhomogeneities of the propagation medium.  

We would note that the diagnostics may be performed using the approximate derivatives directly from equations (\ref{pir},\ref{lar}) on each step of data including. 

\section{Acoustics. Entropy mode and dissipation account. Projecting technique development.}
\label{DA}

In this section we develop results of \cite{LeVer} in direction of \cite{Pere} for 1D homogeneous gas medium. 
The basic system for a viscous and thermo conductive liquid are defined by the momentum,  energy and mass balance. It
 can be written in dimensionless variables based on a
characteristic length of the disturbance, linear 
speed of sound and density, ($c_0$ and  $\rho_0$) such that the
 dimensionless value of the uniform background density $\rho_0$ is equal to
1. It can also be viewed in matrix form as
\begin{equation}\label{bas1}
  \frac{\partial}{\partial t} \psi +L \psi =\tilde{\psi },
\end{equation}
where $L$ is  the  linear matrix  operator
 $$L=\left(
\begin{array}{ccc}
-\delta _{1} D  &1 & 0 \\
1 & -\frac{\gamma \delta _{2}}{\gamma -1}D  &
\frac{\delta_{2}^{} }{\gamma -1} D  \\
1 & 0 & 0
\end{array}
\right)D $$
 with
 $\delta _{1}^{} =\frac{4\mu }{3 \rho_0 c_0\lambda }$, and
 $\delta _{2}^{} =  \frac{\kappa }{ \rho_0 c_0\lambda }\left(\frac{1}{c_v}- \frac{1}{c_p}
\right)$.Here  λ means a characteristic scale of longitudinal perturbation.  (One can note that the expression
$\gamma\frac{\delta_2}{\gamma -1} = - \frac{\kappa}{c_v \rho_0 c_0 \lambda}.$)
 Here $\psi$ is defined as  $\psi (x,t)=\left(
\begin{array}{ccc}
v(x,t) & p(x,t) & \rho (x,t)
\end{array}
\right) ^{T}$   where $v$  represents the $x$-component of the
nondimensional unsteady velocity and $p$ and $\rho$ are nondimensional. The constant parameter of the fluid $ \mu ,
\kappa$ are  viscosity and thermal conductivity respectively. The heat capacities $c_p,c_v$ are normalasid per unit mass, $\gamma=c_p/c_v$.

The formation of these projection operators for the given
one-dimensional flow system gives
\begin{equation}\label{proac}
  P_{1,2} = \left(
\begin{array}{ccc}
\frac{1}{2} \pm \left( \frac{\delta _{2}^{} }{2} -\frac{\beta }{4}
\right)
D& \pm \frac{1}{2} -\frac{\delta _{2}^{} }{2(\gamma
-1)} D& \frac{\delta _{2}^{} }{2(\gamma -1)}
D \\
\pm \frac{1}{2}  & \frac{1}{2} \pm \left( \frac{\beta }{4} -\frac{\gamma
\delta _{2}^{} }{2(\gamma -1)} \right) D& \pm
\frac{\delta _{2}^{} }{2(\gamma -1)} D\\
\pm \frac{1}{2} +\frac{\delta _{2}^{} }{2} D&
\frac{1}{2} \pm \left( \frac{\beta }{4} -\frac{\delta _{2}^{} }{2(\gamma
-1)} \right) D& \pm \frac{\delta _{2}^{} }{2(\gamma
-1)}D
\end{array}
\right)
\end{equation}
define left and right waves $\Pi,\Lambda$. While the third (approximate) projector 
\begin{equation}\label{proheat}
P_{3} =\left(
\begin{array}{ccc}
0 & \frac{\delta _{2}^{} }{\gamma -1} D&
-\frac{\delta
_{2}^{} }{\gamma -1} D\\
0 & 0 & 0 \\
-\delta _{2}^{} D& -1 & 1
\end{array}
\right).
\end{equation}
yields the so-called entropy mode $s$.

In an absorbing fluid, the total energy conservation law
should be considered, the  governing equation  of interest is thus \cite{Pere,Pie}
\begin{equation}\label{I1}
(\rho e +\rho ({\bf v} \cdot {\bf v} )/2)_t+{\nabla }  \cdot {\bf J}
=0
\end{equation}
where,
 ${\bf J} =p{\bf v} +e{\bf v} $
 is the energy flux
density vector,
 $e ,\rho ,{\bf v}$ and  ,$p$
 are internal energy per mass unit, mass density,
velocity, and pressure,  respectively. For ideal gas $e=p/\rho(\gamma-1)$.

The division of the total perturbation field in accordance with mode content and its substitution into \eqref{I1} results in \cite{Pere}
\begin{equation}\label{B}
\frac{\partial E_s}{\partial t}= - div\vec J_a
\end{equation}
where $E_s$ - energy density of entropy mode, $J_a $ density of energy flux for the acoustic waves. Both are results of averaging by period.
So we should take into account the energy losses when a diagnostics is performed.

In one-dimensional case the equation \eqref{I1} reads
\begin{equation}\label{I11}
(\rho e +\rho  v ^2/2)_t+D  J
=0,
\end{equation}
where $ J =p v +e v $.

The operators action splits the system of equations similar to previous sections. The norm and the whole analysis of measurements is more complicated, because of the dissipation originally presented in the problem formulation. Formally, due to the mode evolution equations \cite{Pere}  for $\Pi,\Lambda,s$, the  equality
 \begin{equation}\label{cla}
	[\int_{-\infty}^{\infty}(E_a+E_s)dx]_t=0,
\end{equation}
holds on rapidly decaying functions (localized perturbation), but the integrand is energy perturbation density including the entropy part. A measurementl of the entropy part may be not available, then the balance  \eqref{B} should be taken into account.

Hence the norm choice depends on the measurements access, in the second case the norm
$$  
||\psi||^2=\int_{-\infty}^{\infty}E_adx
 $$
may be used but with its non-conservation account. The resulting algorythm of a diagnostics  in this case depends on a solution form.
 
\section{Conclusion} We presented a 1+1 version of the theory of wave diagnostics. The key tools as it is demonstrated are non trivially generalized, but the basic Banach space and the  functional \eqref{F} are lifted in the prescribed form. 
Account of weak inhomogeneity and dissipation is incorporated by modifcation of correspondent projecting operators, going to nonabelian algebraic approach or including extra (entropy) mode in the second case. The conservation law modification is necessary to introduce the appropriate norm for a wave mode contribution estimation.      
\section{Acknowledgment}
 The work is supported by Ministry of Education and Science of the Russian Federation (Contracts No N GZ 3.1127.2014K)"

\label{lastpage}
\end{document}